\documentstyle[aps,prl,epsf,rotate]{revtex}
\begin{document}

\twocolumn[\hsize\textwidth\columnwidth\hsize\csname     
@twocolumnfalse\endcsname

\title{Transition from an electron solid to the sequence of fractional
quantum Hall states at very low Landau level filling factor}

\author{W. Pan$^{1,2}$, H.L. Stormer$^{3,4}$, D.C. Tsui$^1$, L.N.
Pfeiffer$^4$, K.W. Baldwin$^4$, and K.W. West$^4$}
\address{$^1$Department of Electrical Engineering, Princeton
University,
Princeton, New Jersey 08544}
\address{$^2$National High Magnetic Field Laboratory,
Tallahassee, Florida 32310}
\address{$^3$Department of Physics
and Department of Applied Physics, Columbia University, New
York,
New York 10027}
\address{$^4$Bell Labs, Lucent Technologies, Murray Hill, New
Jersey 07974}

\date{\today}
\maketitle

\begin{abstract}

At low Landau level filling of a two-dimensional electron system,
typically associated with the formation of an electron crystal, we
observe local minima in $R_{xx}$ at filling factors $\nu$ = 2/11, 3/17, 3/19,
2/13, 1/7, 2/15, 2/17, and 1/9. Each of these developing fractional
quantum Hall (FQHE) states appears only above a filling
factor-specific temperature. This can be interpreted as the melting of
an electron crystal and subsequent FQHE liquid formation. The observed
sequence of FQHE states follow the series of composite fermion states
emanating from $\nu=1/6$ and $\nu=1/8$.
\end{abstract}

\vskip0pc]

The electronic state of a two-dimensional electron system (2DES) in
the presence of a high magnetic ($B$) field is the result of a
competition between electrostatics and subtle quantum mechanics,
leading to intricate electron-electron correlations. At extremely high
magnetic field and low carrier concentration electrons are forced on
tiny orbits as compared to their separation. As a consequence mutual
wavefunction overlap vanishes, electrostatics dominates, and an
electron crystal, analogous to a classical Wigner crystal of point
charges, is expected to form \cite{lozovik:jept75,lee:prb79,cfbook1}.
At lower fields, or equivalently
higher electron concentration, wavefunction overlap is appreciable.
The resulting complex electron-electron correlation leads to the
multitude of fractional quantum Hall (FQHE) states \cite{fqhe82,laughlin:prl83,qhebook,fqhebook} at rational
values of Landau level filling factors $\nu=n/s$, where $n$ is the electron
density and $s$ is the magnetic flux density, $s=B/\Phi_0$ with the flux
quantum $\Phi_0=h/e$. The composite fermion (CF) picture \cite{jain:prl89,hlr:prb93,willett:ap97,cfbook2} has been
very successful in classifying these FQHE states into series of
fractions emanating from neighboring even-denominator CF liquids. The
positive identification of the Wigner crystal remains somewhat
controversial, although many different experiments detect behavior
that is most naturally explained in terms of such a phase \cite{cfbook1}. In
particular, in electronic transport experiments the transition to very
high resistances is often equated with the onset of electron crystal
formation. Such a solid phase is easily pinned by residual potential
fluctuations in contrast to the quantum liquids of the FQHE, which can
circumvent such obstacles. 
The weak pinning in the solid phase has been observed in recent microwave
measurements \cite{engel:ssc97}.
 
The transition point from the series of FQHE states to the Wigner
crystal is one of the oldest questions in the field. Laughlin's
original publication on the FQHE estimated this transition to
occur around $\nu \sim 1/10$ \cite{laughlin:prl83} 
and subsequent calculations refined it 
to $\nu \sim 1/6.5$ 
\cite{lam:prb84,levesque:prb84,chui:prb90,zhu:prb95,yang:prb01}.
Experiment shows a rather more
complex behavior. Indeed, even the highest quality specimens -- least
subject to distortions and hence spurious results
from glassy behavior -- show at low temperatures a rapid divergence of
the resistance for magnetic field values higher than $\nu=1/5$. However,
even above $\nu=1/5$ the resistance diverges
in a narrow window $2/9 \le \nu \le 1/5$ \cite{jiang:prl90,du:ssc96}. 
This reentrant behavior
is interpreted as resulting from complicated, multiple intersections
of the non-monotonic ground state energy curve of the FQHE liquid with
the smooth equivalent curve for the electron crystal as a function of
$\nu$. Also, below $\nu=1/5$ -- i.e. for higher magnetic fields -- the
situation is more complex than may be deduced from a single transition
point. Weak local minima in the resistance $R_{xx}$ have been observed at
filling factors $\nu=1/7$ \cite{goldman:prl88} 
and $\nu=2/11$ \cite{mallett:prb88} hinting at the existence
of FQHE states in the ``forbidden'' Wigner crystal regime, although such
features always disappear as $T \to 0$. Furthermore, optical experiments
\cite{buhmann:prl,hayne:ss92}, 
performed at elevated temperatures of $\sim$ 500~mK, observe
features in the luminescence lines at filling factors $\nu=1/7$ and
$\nu=1/9$. Initially, this has been taken as additional reentrances of
the FQHE liquid into the Wigner crystal phase at $\nu=1/7$ and 1/9.
Later theoretical work \cite{price} interprets the data as those FQHE states
representing a higher temperature liquid phase, while the Wigner
crystal remains the $T=0$ ground state at $\nu=1/7$ and $\nu=1/9$. This points
again to a much more complex interrelation between Wigner crystal and
the FQHE states.

In this paper we present electronic transport data on the ultra-high
mobility two-dimensional electron system in a GaAs/AlGaAs quantum well
and observe local minima in $R_{xx}$ at a multitude of low Landau level
filling factors $\nu=2/11$, 3/17, 3/19, 2/13, 1/7, 2/15, 2/17, and 1/9.
Each of these developing FQHE states appears only above a filling
factor-specific temperature. The FQHE sequence follows the series of
composite fermion (CF) Landau levels expected to emanate from a CF
liquid at $\nu=1/6$ and $\nu=1/8$. These observations demonstrate the
continued applicability of the CF model for the FQHE states well into
the Wigner crystal regime. The series of FQHE states only exist at
elevated temperatures, probably above the point at which the solid has
melted. 
 
The quantum well is 500\AA~~wide and $\delta$-doped symmetrically from
both sides at a distance of 2200\AA. The electron density of
$n \sim 1.0 \times 10^{11}$~cm$^{-2}$ and mobility of 
$\mu \sim 10 \times 10^6$~cm$^2$/Vsec are established
after illumination by a red light emitting diode.
At this density, only one electric subband is occupied. The
specimen has a size of 5~mm $\times$ 5~mm, with 8 indium contacts
around its perimeter. The data were taken in
dilution refrigerators with different base temperatures placed in
three different magnets: $\sim$ 35~mK in 18~Tesla, $\sim$ 70~mK in 33~Tesla, and
$\sim$ 70~mK in 42~Tesla. Low frequency ($\sim$ 3-4~Hz) lock-in amplifier
techniques were employed to measure $R_{xx}$, with an excitation current of
0.1 - 1~nA to avoid electron heating.

The high quality of this sample (Fig. 1) is evident from the FQHE
sequences $p/(2p\pm 1)$ showing $R_{xx}$ features up to $\nu=10/19$ and 10/21,
the sequences $p/(4p\pm 1)$ showing features up to $\nu=6/23$ and 6/25, and
the observation of novel FQHE states, {\it e.g.} at $\nu$ = 4/11, 
between $\nu=1/3$ and $\nu=2/5$. Most
remarkable is the flat background around $\nu=1/4$. All previous quantum
Hall samples, show a rising background starting around $\nu=2/7$,
indicating that the sample is approaching the insulating regime. No
such background is apparent here and $R_{xx}$ at $\nu=1/4$ is practically
temperature independent as expected for CF liquid at
an even-denominator filling factor. This extremely high quality 2DES
allows us to pursue our transport measurement to very low filling
factors.

Fig. 2 shows $R_{xx}$ at different temperatures for filling factors $1/5
\ge \nu \ge 1/7$. In all traces the $\nu=1/5$ state is well developed,
showing vanishing resistance in $R_{xx}$ and a clean quantized Hall plateau
in $R_{xy}$ (not shown). Similar to all high-quality samples, as
$T\to 0$, $R_{xx}$ diverges for $\nu<1/5$ and for 
$2/9<\nu<1/5$ (just to the left
off the graph). The 80~mK trace clearly shows this divergent behavior.
In fact, $R_{xx}$ at $\nu \sim 0.21$ shows 
activated behavior with a characteristic
energy of 1.1~K. This value is very close to those obtained
earlier in samples of similar density but with poorer FQHE features
and a rising background around $\nu=1/4$ \cite{jiang:prl90,du:ssc96}. 
This seems to
suggest that a limiting state has been reached. The re-entrant
insulating phase and, by extension, the insulating phase beyond $\nu=1/5$
appear to be of intrinsic origin and no longer dominated by
appreciable disorder. These data are taken as indicating the formation
of a pinned Wigner solid ground state for filling factors $\nu<2/9$,
interrupted solely by a FQHE state at $\nu=1/5$. In particular, for
$\nu<1/5$ the 2DES becomes insulating and no further FQHE features are
observed for $T \to 0$. However, raising the temperature to 115~mK through
165~mK uncovers multiple minima in $R_{xx}$ that occur at distinct, rational
fractional filling factors $\nu=2/11$, 3/17, 3/19, 2/13, and 1/7. They
represent the $p/(6p \pm 1)$ series of FQHE states, emanating from the CF
liquid at $\nu=1/6$. The observation of this sequence is a demonstration
of the applicability of the CF model even to these very low filling
factor FQHE states within the regime of the Wigner solid, albeit at
elevated temperatures. Jumping ahead to the data of Figure 3, taken at
the yet more extreme conditions of a hybrid magnet, one can even
discern features at $\nu=2/15$, 2/17 and 1/9. These are the first
representatives of the $p/(8p \pm 1)$ sequence associated with the CF
liquid state at $\nu=1/8$. Both observations attest to the wide
applicability of the CF classification scheme. 

Returning to Fig.~2 one notices that the resistance values are high
compared to typical $R_{xx}$ values in the FQHE regime. 
They are in the M$\Omega$
range, rather than the usual k$\Omega$ range (see Fig.~1). For any
temperature, as the magnetic field is raised, $R_{xx}$ becomes increasingly
noisy and ultimately collapses \cite{willett:prb88}, 
taking on even negative values as
illustrated in the 135~mK trace. This is accompanied by a sharply
increasing out-of-phase component, indicating resistances in the range
of $R \sim 1/(\omega$C) $\sim$ 1~G$\Omega$ 
(using $\sim$ 3~Hz and $\sim$ 100~pF stray capacitance). However,
$R_{xx}$ is well behaved up to the turnaround point 
({\it  e.g.} 27~Tesla at 135~mK)
for all temperatures measured.
We limit our analysis to this regime. 

There seems to be an optimum temperature for the development of any
given fraction and a lowest and highest temperature cut-off for its
observation. The high-$T$ limit, as usual, is a measure for its energy
gap. At temperatures comparable with this gap the number of thermally
excited quasi-particles becomes appreciable and flood the
characteristic transport features of the FQHE. The existence of a
low-$T$ limit is unusual, is not observed for higher filling factors,
and characteristic of the Wigner crystal regime. The inset to Fig.~2
summarizes these limits as determined in a qualitative manner from
many traces such as seen in Fig.~2. As expected, the high-$T$ limit is
maximum at $\nu=1/5$ and $\nu=1/7$ and drops towards the center. This
reflects the decreasing energy gap of the FQHE states in the CF model
as the CF liquid at $\nu=1/6$ is approached. 

In contrast to the high-$T$ limits, the low-$T$ limits show a monotonic
increase with increasing $1/\nu$ (higher $B$-field). The origin of this
transport behavior at such low filling factors remains unclear, but is
most probably intimately related to electron solid formation. The most
likely interpretation of a low-$T$ limit for FHQE observation in Fig.~2
involves a two-phase picture in this low-filling factor regime and,
with increasing temperature, a transition from a Wigner crystal to an
electron liquid, which assumes FQHE correlation at the relevant
filling factors. The Wigner crystal is the $T=0$ ground state, whereas
the different fractions of the FQHE form the ground state at elevated
temperatures. This progression seems to be at odds with a simple
thermodynamic argument \cite{price}: The ground state is determined by the free
energy, $F=E-TS$, where $E$ is the energy at $T=0$ and $S$ is the entropy.
If the Wigner solid 
is the ground state at $T=0$ it is expected to
remain the ground state for non-zero $T$, since its excitations are
gapless and hence $TS$ grows rapidly compared to the case of a FQHE
state, which is gapped. While this argument always holds for
infinitesimal temperatures Price {\it et al.} \cite{price} 
have shown that it fails as $T$
becomes a substantial fraction of the FQHE gap energy. At this stage
the very high density of states for excitations at the edge of the
FQHE gap gives rise to an exponential growth in $TS$ and wins out over
$TS$ of the Wigner solid, which rises only as a power law. Hence the
magnitude of the energy gap is the decisive parameter for this phase
transition rather than any small difference in the ground state
energies between solid and liquid. 

The calculated transition temperature for the $\nu=1/7$ state is $\sim$
600~mK
for a clean sample with parameters close to those of our specimen
\cite{price}.
This value drops to $\sim$ 400~mK when disorder is taken into account
phenomenologically and approximate FQHE gap energies are
inferred from experiment. Both values considerably exceed the low-$T$
limit of $\sim$ 135~mK for $\nu=1/7$. However, given
our very limited ability to treat disorder and its impact on transport
in these highly correlated phases and given our present inability to
determine experimentally the $\nu=1/7$ energy gap -- not to mention
the magneto-roton gaps, which set the energy scale for the entropy -- an
agreement within a factor of three may be quite satisfactory. In
particular, the trend of increasing disorder moving the transition to
lower temperatures allows for a simple rationale to account for the
discrepancy.

On the other hand, the general features of the inset of Fig.~2 seem
not to be consistent with the essence of the melting calculations.
According to the model, melting occurs when $T$ has reached a
substantial fraction of the FQHE gap energy. Taking the high-$T$ limit
of the inset to Fig.~2 as a measure for the energy gap of the FQHE
liquid one would expect the lower limit -- taken to reflect the
melting temperature - to track the former. This is definitely not the
case. In fact, both seem to be rather independent from each other: the
``gaps'' (high-$T$ limit) show the characteristic minimum around $\nu=1/6$,
whereas the ``melting temperature'' (low-$T$ limit) rises monotonically
with inverse $\nu$.
Calculations for higher-denominator
states than 1/7 (and 1/9) may well reveal such a dependence, but, it
appears, that a more general consideration may be more fruitful in
uncovering the origin of the general trend seen in the inset to Fig.~2.

A simple classical Wigner crystal melting picture \cite{grimes:prl79}, 
in which the
electrons are viewed as point charges on a triangular lattice yields a
melting temperature of $T_m \sim 260$mK for our sample.
This temperature is in reasonable proximity to the
experimental overall ``melting temperature'' of $\sim$ 120~mK of the inset of
Fig.~2. Of course, this crude calculation yields a filling
factor-independent value (relatively well reflected in experiment for
$1/\nu \ge 5.5$) since it is independent of $B$-field, which
determines the ``size'' of the electrons via the magnetic length.
Eventually, around $\nu \ge 1/5$ this simple model fails totally and
quantum correlations dominate, leading to the ascendance of the FQHE
states. 

Independent of the model used to estimate the transition temperature,
melting of the electron crystal and subsequent formation of series of
FQHE liquids at higher temperatures is able to describe the
qualitative features of $R_{xx}$ in our experiments. For a quantitative
comparison between experiment and theory much remains to be achieved
and, most likely, yet higher quality 2DES specimens are required. The
exceedingly high resistances of Fig.~2 are probably an indicator for
the influence of remnant disorder and density fluctuations. Once the
Wigner crystal has melted, one would expect for the developing FQHE
states resistances in the range of k$\Omega$'s, rather than the M$\Omega$'s,
observed in Fig.~2. These high resistance values are probably due to
density fluctuations leading to coexisting patches of electron
crystal and electron liquid at any given $B$-field. The percolating
electron path through this patchwork would make electron transport
considerably more difficult. Alternatively, these high resistances may
be of an intrinsic origin, in which the transition from the solid to
the liquid is more complex and states at
rational fractional fillings evolve gradually, carrying with them
aspects of the solid that makes them more susceptible to localization.

In summary, in an exceptionally high quality sample, $R_{xx}$ minima arise
at Landau level filling factors $\nu=2/11$, 3/17, 3/19, 2/13, 1/7, 2/15,
2/17,
and 1/9 as clear signatures of developing FQHE states. Each FQHE state
appears only above a filling factor-specific temperature and in a
certain temperature range. This behavior is well accounted for by
melting of the Wigner crystal phase, expected to exist at such low
filling factors, into the series of FQHE liquids prescribed by the
composite fermion model. 

We would like to acknowledge the staff of the
NHMFL, especially E. Palm, T. Murphy, J. Pucci, and R.
Smith for experimental assistance. We thank L.W. Engel, H. A. Fertig,
J.K. Jain, and T. Knuuttila for helpful discussion. A portion of this work was
performed at the NHMFL, which is supported by NSF Cooperative
Agreement No. DMR-9527035 and by the State of Florida. D.C.T. and W.P.
are supported by the AFOSR, the DOE, and the NSF.  H.L.S. is supported
by DOE and the W.M. Keck Foundation.

\begin{figure}
\caption{
Diagonal resistance $R_{xx}$ of a sample of 
n = $1.0 \times 10^{11}$~cm$^{-2}$ and 
$\mu =10 \times 10^6$~cm$^2$/Vsec. Arrows mark several key
fractional Landau level filling factors.
}
\end{figure}

\begin{figure}
\caption{
$R_{xx}$ above 20~T at various temperatures. The vertical, dashed lines show
the positions of the Landau level filling factors $\nu=1/5$, 2/11, 3/17,
3/19, 2/13, and 1/7. The inset summarizes high-$T$ limits (open squares)
and low-$T$ limits (solid dots) for the observation of features in $R_{xx}$
at various $\nu$. The dashed line is only a guide to the eye. The
low-T limits may be viewed as the melting line from Wigner crystal to
FQHE liquids and the high-T limits are measures the energy gap of FQHE
states (see text for details).
}
\end{figure}

\begin{figure}
\caption{
$R_{xx}$ of the same sample as in Fig.~2 placed in a hybrid magnet with
fields up to 42~Tesla at two different temperatures.
}
\end{figure}


\begin{references}

\bibitem{lozovik:jept75}
Y.E. Lozovik and V.I. Yudson, JEPT Lett. {\bf 22}, 11 (1975).

\bibitem{lee:prb79}
H. Fukuyama and P.A. Lee, Phys. Rev. B {\bf 18}, 6245 (1979).

\bibitem{cfbook1}
For a review of recent theoretical and experimental results on Wigner
crystal, see, for example, the chapters by H.A. Fertig and M. Shayegan
in {\it Perspectives in Quantum Hall Effect}, S. Das Sarma and A. Pinczuk
(Eds.), Wiley, New York (1996), and references therein.

\bibitem{fqhe82}
D.C. Tsui, H.L. Stormer, and A.C. Gossard, Phys. Rev. Lett. {\bf 48},
1559 (1982).

\bibitem{laughlin:prl83}
R.B. Laughlin, Phys. Rev. Lett. {\bf 50}, 1395 (1983).

\bibitem{qhebook}
{\it The Quantum Hall Effect}, R.E. Prange and S.M. Girvin (Eds.),
Springer, New York (1990).

\bibitem{fqhebook}
{\it The Quantum Hall Effects}, T. Chakraborty and P. Pietilainen, Springer,
New York (1995).

\bibitem{jain:prl89} 
J.K. Jain, Phys. Rev. Lett. {\bf 63}, 199 (1989).

\bibitem{hlr:prb93}
B.I. Halperin, P.A. Lee, and N.Read, Phys. Rev. B {\bf 47}, 7312 (1993).

\bibitem{willett:ap97}
R.L. Willett, Adv. Phys. {\bf 46}, 447 (1997).

\bibitem{cfbook2}
{\it Composite Fermions: A Unified View of the Quantum Hall Regime}, O.
Heinonen (Edt), World Scientific, Singapore (1998).

\bibitem{engel:ssc97}
L.W. Engel, C.C. Li, D. Shahar, D.C. Tsui, and M. Shayegan, Solid
State Commun. {\bf 104}, 167 (1997).

\bibitem{lam:prb84}
P.K. Lam and S.M. Girvin, Phys. Rev. B {\bf 30}, 473 (1984).

\bibitem{levesque:prb84}
D. Levesque, J.J. Weis, and A.H. MacDonald, Phys. Rev. B {\bf 30}, 1056
(1984).

\bibitem{chui:prb90}
K. Esfarjani and S.T. Chui, Phys. Rev. B {\bf 42}, 10758 (1990).

\bibitem{zhu:prb95}
X. Zhu and S.G. Louie, Phys. Rev. B {\bf 52}, 5863 (1995).

\bibitem{yang:prb01}
K. Yang, F.D.M. Haldane, and E.H. Rezayi, Phys. Rev. B {\bf 64}, 08~1301
(2001).

\bibitem{jiang:prl90}
H.W. Jiang, R.L. Willett, H.L. Stormer, D.C. Tsui, L.N. Pfeiffer, and
K.W. West, Phys. Rev. Lett. {\bf 65}, 633 (1990).

\bibitem{du:ssc96}
R.R. Du, H.L. Stormer, D.C. Tsui, L.N. Pfeiffer, K.W. Baldwin, and
K.W. West, Solid State Commun. {\bf 70}, 2944 (1996).

\bibitem{goldman:prl88}
V.J. Goldman, M. Shayegan, and D.C. Tsui, Phys. Rev. Lett. {\bf 61}, 881
(1988).

\bibitem{mallett:prb88}
J.R. Mallett, R.G. Clark, R.J. Nicholas, R.L. Willett, J.J. Harris,
and C.T. Foxon, Phys. Rev. B {\bf 38}, 2200 (1988).

\bibitem{buhmann:prl} 
H. Buhmann, W. Joss, K.v. Klitzing, I.V. Kukushkin, A.S. Plaut, G.
Martinez, K. Ploog, and V.B. Timofeev, Phys. Rev. Lett. {\bf 65}, 1056
(1990); {\it ibid.} {\bf 66}, 926 (1991).

\bibitem{hayne:ss92}
M. Hayne, M.K. Ellis, A.S. Plaut, A. Usher, and K. Ploog, Surface
Science {\bf 263}, 39 (1992). 

\bibitem{price}
R. Price, X. Zhu, P.M. Platzman, and S.G. Louie, Phys. Rev. B {\bf 48},
11473 (1993); P.M. Platzman and R. Price, Phys. Rev. Lett. {\bf 70}, 3487
(1993).

\bibitem{willett:prb88}
R.L. Willett, H.L. Stormer, D.C. Tsui, L.N. Pfeiffer, K.W. West, and
K.W. Baldwin, Phys. Rev. B {\bf 38}, 7881 (1988).

\bibitem{grimes:prl79}
C.C. Grimes and G. Adams, Phys. Rev. Lett. {\bf 42}, 795 (1979).


\end{references}
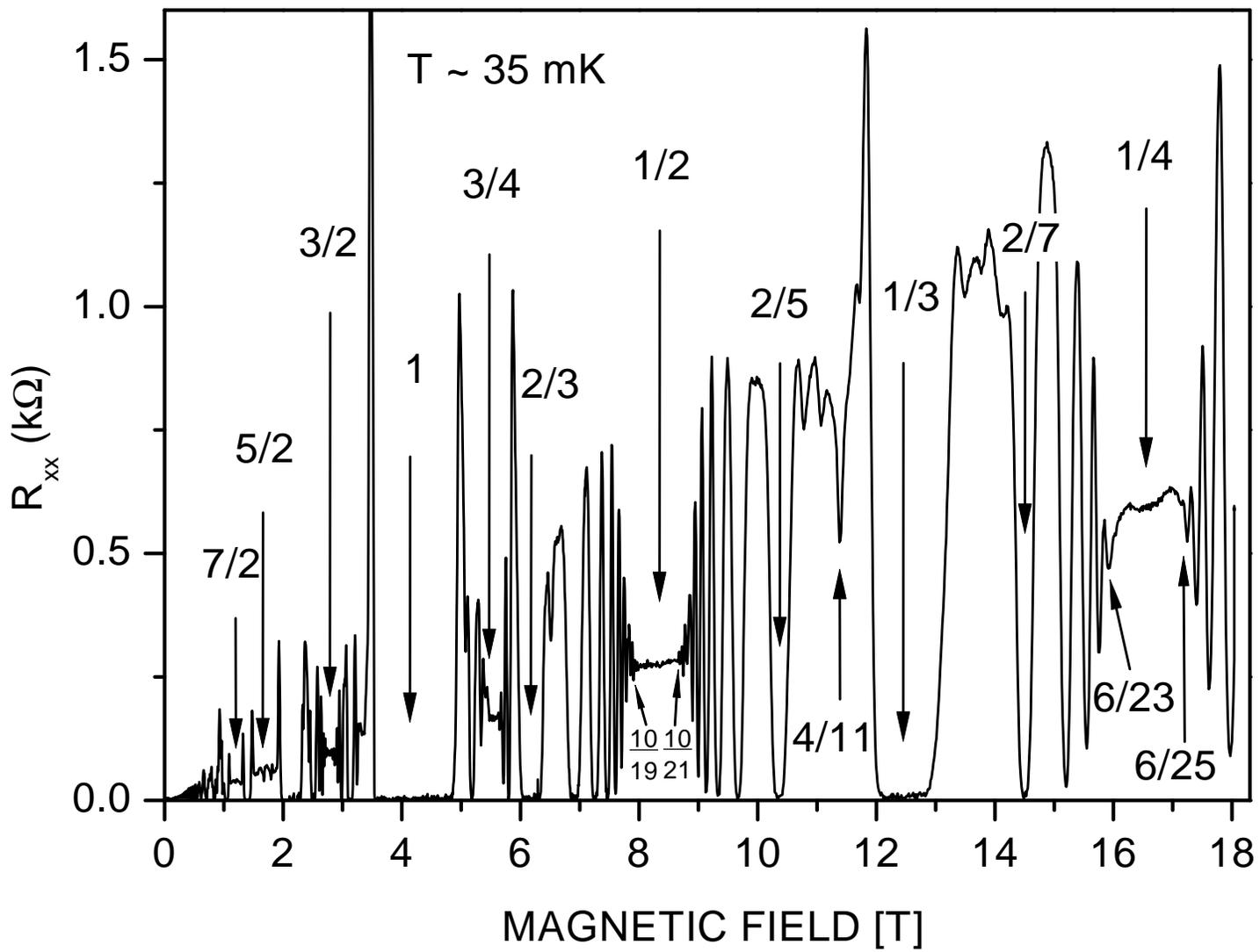
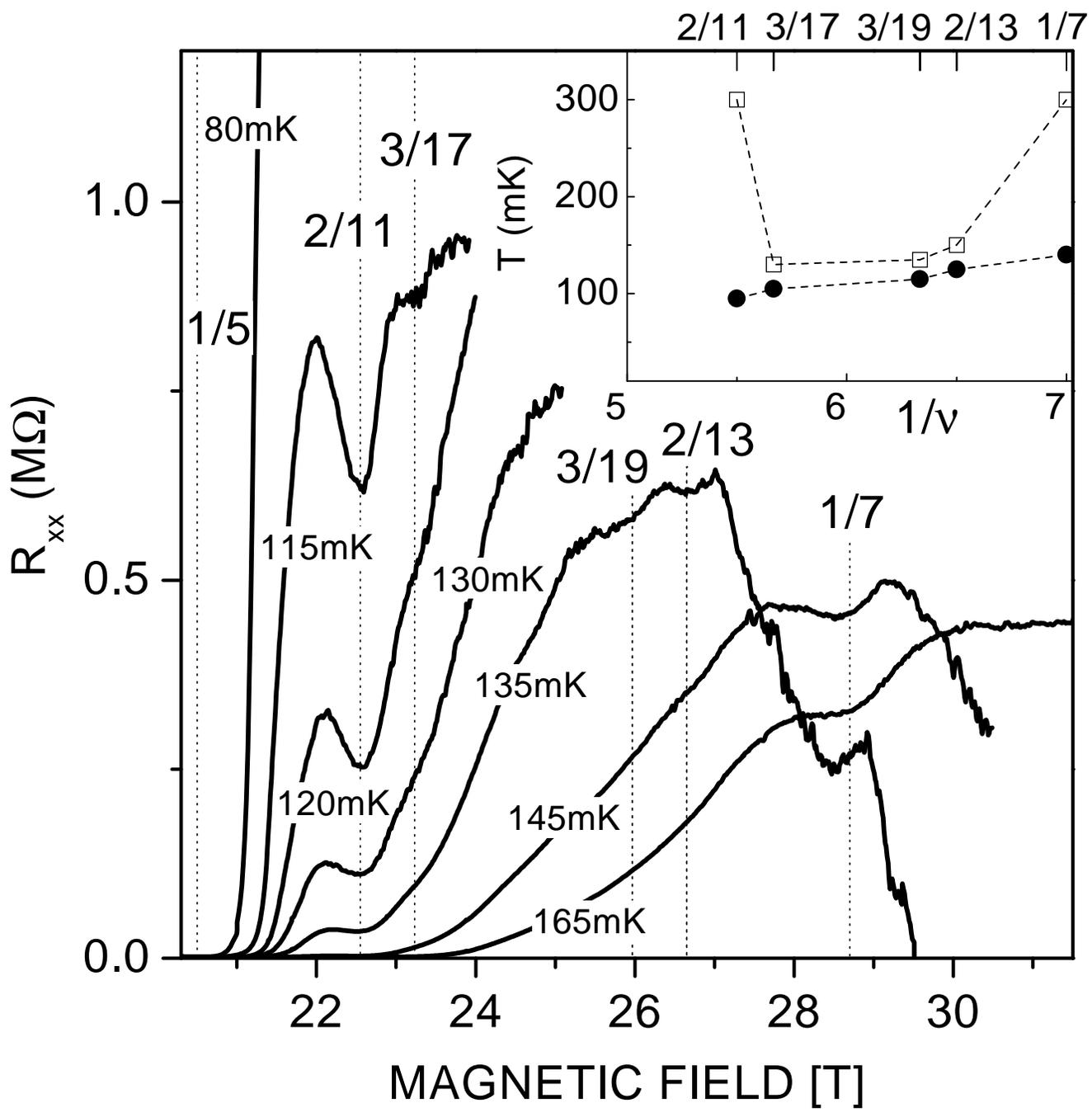
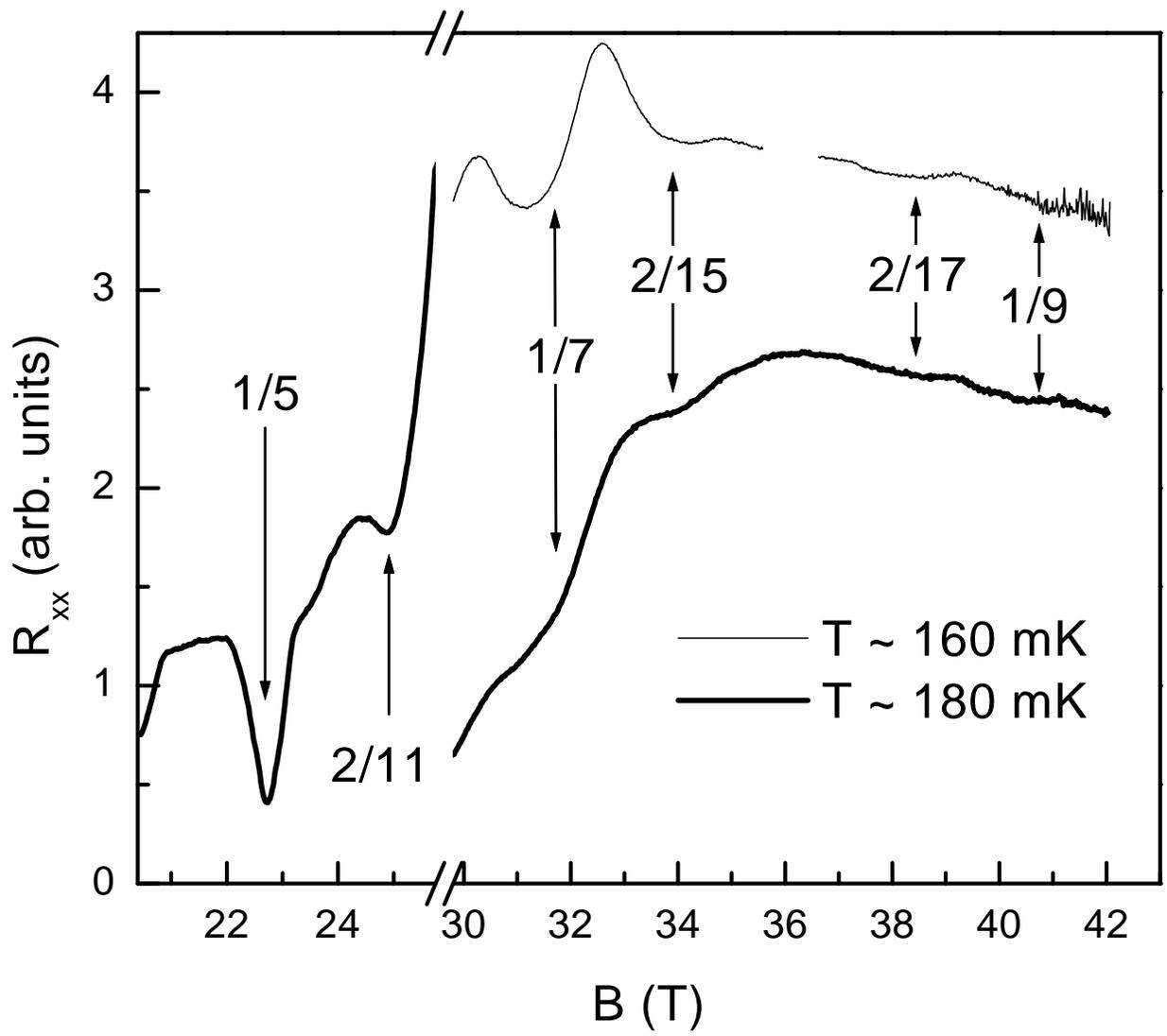
\end{document}